\documentstyle[preprint,aps,12pt]{revtex}

\def\half{{\textstyle{1\over2}}}
\def\third{{\textstyle{1\over3}}}

\def\beq{\begin{equation}}
\def\eeq{\begin{equation}}

\def\pmb#1{\setbox0=\hbox{$#1$}%
\kern-.025em\copy0\kern-\wd0
\kern.05em\copy0\kern-\wd0
\kern-.025em\raise.0433em\box0}

\let\varkappa\kappa
\skip\footins = 14pt 
\newif\ifMarginNotes \MarginNotesfalse

\def\mrgn#1{\ifMarginNotes\setbox0=\vtop{\hsize 10pc
   {\footnotesize\noindent\relax #1\par}}\leavevmode
   \vadjust{\dimen0=\dp0 \dimen1=\ht0\advance\dimen1 by .5ex
  \advance\dimen0 by -.5ex
  \kern-\dimen1\hbox{\kern\hsize\kern.5pc$\leftarrow$
  \box0}\kern-\dimen0}\fi}

\ifMarginNotes
\global\def\Mrgnpar#1{\marginpar[\raggedright\footnotesize
    $\rightarrow$\enspace#1]{\raggedright\footnotesize$\leftarrow$\enspace#1}}%
\global\def\Mrgnparb#1{\marginpar[\raggedright\footnotesize#1\enspace
    $\searrow$]{\raggedright\footnotesize$\swarrow$\enspace#1}}%
\else
\global\def\Mrgnpar#1{\relax}%
\global\def\Mrgnparb#1{\relax}%
\fi

\begin{document}
\tighten

\nonfrenchspacing
\flushbottom
\title{Anyons and Chiral Solitons on a Line\footnotemark[1]}

\footnotetext[1] {\baselineskip=16pt This work is supported in part by funds
provided by  the U.S.~Department of Energy (D.O.E.) under contracts
\#DE-FC02-94ER40818, \#DE-FG02-91ER40676, and by INFN (Istituto Nazionale di Fisica
Nucle\-are, Frascati, Italy).  \hfil \break 
\hfil MIT-CTP-2541 \hfil  June 1996  \hfil BU\# xxxxxx \qquad  \break
}

\author{U. Aglietti,$^\dagger$ L. Griguolo,$^\dagger$ R.~Jackiw,$^\dagger$ S.-Y.
Pi,$^\ddagger$ \\ and \\
D. Seminara$^\dagger$\\\bigskip}

\address{$^\dagger$Center for Theoretical Physics\\ Massachusetts Institute of
Technology\\ Cambridge, MA ~02139--4307\\
and\\
$^\ddagger$ Department of Physics \\ Boston University \\
Boston, MA ~02215}

\maketitle
\begin{abstract}%
We show that excitations in a recently proposed gauge theory for anyons on a line in fact
do not obey anomalous statistics. On the other hand, the theory supports novel chiral
solitons. Also we construct a field-theoretic description of lineal anyons, but gauge fields
play no role. 
\vspace{.5in}

\centerline{Submitted to {\it Physical Review Letters}.}
\end{abstract}

\setcounter{page}{0}
\thispagestyle{empty}

\newpage

A (nonrelativistic) field theoretic description for the quantum mechanics of planar
particles with unconventional statistics -- anyons -- makes use of a Chern-Simons gauge
field in a second quantized formalism. Recently there has appeared in these pages an
article offering a similar description for particles on a line \cite{ref:1}. However, the
claimed results are incorrect; apparently inattention to signs has led to error.

In this Letter, we analyze the model of Ref.\cite{ref:1}. We present an alternative
nonrelativistic field theory, which succeeds in describing lineal anyons, but is not a
gauge theory.  Finally we show that the model of Ref.\cite{ref:1}, though failing to achieve
its announced goal, possesses  an interesting and novel soliton structure.
\bigskip

{\bf A.}\quad A nonrelativistic gauge field theory that leads to planar anyons is the
nonlinear Schr\"odinger equation, gauged by a Chern-Simons field and governed by the
Lagrange density
 \begin{eqnarray}
 {\cal L}_{(2+1)} &=& \frac1{4\bar\varkappa}\epsilon^{\alpha\beta\gamma} A_\alpha
F_{\beta\gamma}
 + i\hbar \Psi^* (\partial_t + iA_0)\Psi - 
\frac{\hbar^2}{2m} \sum_{i=1}^2 \left| (\partial_i + iA_i)\Psi\right|^2
- V(\rho) \label{eq:1}\\
\rho &\equiv& \Psi^*\Psi .\nonumber 
 \end{eqnarray}
Here $\Psi$ is the Schr\"odinger quantum field, giving rise to charged bosonic particles
after (second) quantization. $A_\mu$ possesses no propagating degrees of freedom; it
can be eliminated leaving a statistical Aharanov-Bohm interaction  between the
particles. $V$ describes possible nonlinear self-interactions, {\it e.g.}, $V(\rho) \propto
\rho^2$ for the cubic Schr\"odinger equation. 

When analyzing the lineal problem, it is natural to consider a dimensional reduction of
(\ref{eq:1}), by suppressing dependence on the second spatial coordinate and renaming
$A_2$ as $(mc/\hbar^2) B$. In this way one is led to a $B$--$F$ gauge theory,
described by the Lagrange density 
 \begin{equation} 
 {\cal L}_{(1+1)} = \frac1{2\varkappa} B \epsilon^{\mu\nu} F_{\mu\nu}
+ i\hbar \Psi^*(\partial_t + iA_0)\Psi - 
\frac{\hbar^2}{2m} \left| (\partial_x + iA_x)\Psi\right|^2
- \frac{mc^2}{2\hbar^2} B^2 \rho - V(\rho) .\label{eq:2}
\end{equation}
Here, $\varkappa \equiv (\hbar^2/mc)\bar\varkappa$ is dimensionless. (We retain
physical constants: $c$ is the light velocity, which plays no role in the following.) 
Eliminating the
$B$ and
$A_\mu$ fields decouples them completely, in the sense that the phase of $\Psi$ may be
adjusted so that the interactions of the $\Psi$ field are solely determined by  $V$, and
particle statistics remain unaffected.

$\Bigl[$We observe here an interesting pattern:   dimensional
reduction  of ${\cal L}_{(2+1)}$ in (\ref{eq:1}) and $V \propto
\rho^2$,  with respect to {\it space},
 results in a completely integrable system on
$(1+1)$-dimensional space-time: the nonlinear, cubic Schr\"odinger equation.
On the other hand, reduction  with respect to {\it time\/} results in a completely
integrable system in two spatial dimensions (provided the cubic nonlinearity is of
definite strength): the Liouville equation \cite{ref:2}.$\Bigr]$

In order to make the vector potential $A_\mu$ and the $B$ field dynamically active,
thereby allowing the $\Psi$ particles to interact even in the absence of $V$, we include a
kinetic term for $B$, which could be taken in the Klein-Gordon form. However, we prefer
a simpler expression that describes ``chiral" Bose fields, propagating only in one
direction, 
whose Lagrangian density is proportional to
 $\pm \dot B B' - v B' B'$ \cite{ref:3}. (Dot/prime indicate
differentiation with respect to time/space.) Here $v$ is a velocity and the consequent
equations of motion for this kinetic term (without further interaction) are solved by
$B=B(x\pm vt)$ (with suitable boundary conditions at spatial infinity), describing
propagation with velocity
$\mp v$.  Note that $\dot B B'$ is not invariant against a Galileo transformation, which is
a symmetry of ${\cal L}_{(1+1)}$ and of $B' B'$: 
performing a Galileo boost on $\dot B B'$ with velocity $\tilde v$ gives rise to 
$\tilde v B' B'$,
 effectively boosting the $v$~parameter by
$\tilde v$. Consequently one can drop the $ v B'B'$ contribution to the kinetic $B$
Lagrangian, thereby selecting to work in a global 
 ``rest frame". Boosting a solution in this rest frame then produces a solution to the theory
with a $B'B'$ term. 

In view of the above, we choose the $B$-kinetic Lagrange density to be
 \begin{equation}
 {\cal L}_B = \pm \frac1{\hbar} \dot B B' \label{eq:3}
\end{equation}
and the total Lagrange density is 
 \begin{equation}
 {\cal L} = {\cal L}_B + {\cal L}_{(1+1)}. \label{eq:4}
\end{equation}

It is still possible to remove the $A_\mu$ and $B$ fields by a Hamiltonian reduction as
described in Ref.\cite{ref:3a}, and by phase redefinition of $\Psi$. Once this is done, one is
left with the Lagrangian
 \begin{eqnarray}
 L &=& \int {\rm d}x\, i\hbar \Psi^* \dot\Psi - H \label{eq:5}\\
 H &=& \frac{\hbar^2}{2m}\int{\rm d}x   \Pi^* \Pi  \label{eq:6}\\
 \Pi &\equiv& (\partial_x \pm i\varkappa^2\rho) \Psi .  \label{eq:7}
\end{eqnarray}
For simplicity, $V$ has been omitted.

Quantization is straightforward: $\Psi$ and $\Psi^*$ are conjugates; when the
Hamiltonian is taken in the form (\ref{eq:6}), which is Hermitian but {\it not\/} normal
ordered, one derives the following Schr\"odinger equation for the two-body wave
function $\psi(t; x_1, x_2)$:
\begin{mathletters}\label{eq:allof8}
\begin{equation}
\psi(t; x_1, x_2)  \equiv \frac1{\sqrt2} \langle 0 \mid \Psi(t, x_1)\Psi(t, x_2) \mid 2
\rangle
\end{equation}
\begin{equation}
i\hbar \partial_t\psi = \frac{\hbar^2}{2m}\Bigl\{
  \Bigl(\frac{1}{i}\partial_{x_1} \pm \varkappa^2 \delta(x_1-x_2)\Bigr)^2 +
   \Bigl(\frac{1}{i}\partial_{x_2} \pm \varkappa^2 \delta(x_2-x_1)\Bigr)^2
\Bigr\}\psi . \label{eq:8b}
\end{equation}
\end{mathletters}%
\indent
This is the theory presented in Ref.\cite{ref:1}. It is in the subsequent analysis that the
author goes wrong: he claims that the interaction with the ``vector potential" $\pm
\hbar\varkappa^2 \delta (x_1-x_2)$ can be removed by redefining the wave function
$\psi$ with a step-function phase $\psi =\tilde\psi e^{\mp i\varkappa^2\theta(x_1-x_2)}$.
But this is incorrect: a phase redefinition can remove the potential from the ``1" kinetic
term or from the ``2" kinetic term, but not from both! Indeed removing the $\delta$
function in one term inserts it with the {\it same\/} sign in the other term. This is as it
should be: we have already remarked that the theory is not Galileo invariant. More
specifically, a two-body vector potential is Galileo invariant only if it is an odd function
\cite{ref:4}, while the
$\delta$ function in (\ref{eq:allof8}) is even. Certainly one cannot
transform a Galileo-noninvariant equation to a noninteracting, Galileo-invariant one.

We may solve Eq.(\ref{eq:allof8}). The theory is time-  and space-translation invariant,
so one can separate the time and center-of-mass coordinates
\begin{equation}
\psi(t; x_1, x_2) = e^{-i({Et}/{\hbar})} e^{i({P}/{\hbar}) {(x_1+x_2)}/{2}}
u(x_1-x_2) . \label{eq:9}
\end{equation}

The wave function for relative motion satisfies
\begin{equation}
\frac1m\left[-\hbar^2\partial^2_x +\Bigl(\frac P2 \pm \hbar\varkappa^2\delta(x)\Bigr)^2
\right] u(x) = Eu(x) . \label{eq:10}
\end{equation}
The presence of the total momentum $P$ vividly demonstrates the absence of Galileo
invariance. Explicit solution requires regulating the $\delta$ function, to give meaning
to  $\delta(x)\delta(x)$. Once this is done, one finds a free odd solution
\begin{mathletters}
\label{eq:allof11}
\begin{equation}
u_-(x) = \sin k x\qquad \hbar^2 k^2=mE - P^2/4 \ge 0
\end{equation}
while the even solution reads
\begin{equation}
u_+ (x) = \sin  k \left| x\right| 
\end{equation}
\end{mathletters}%
{\it i.e.}, there is total reflection with $\pi$ phase shift, and no transmission. 
 Evidently
the odd solution sees no potential, while the even one moves in an effective potential
\begin{equation}
V_{\rm effective} (x) =
\frac{P^2}{4m} + \frac{2\hbar^2}{m} \frac{\delta(x)}{\left|x\right|}
\label{eq:12}
\end{equation}
Note that the coupling strength $\varkappa$ has disappeared.  Alternatively, we may
recognize the wave functions (\ref{eq:allof11}) as  solutions to the Schr\"odinger equation
with a $\delta$-function potential, in the limit that its strength becomes infinite. Both from
(\ref{eq:10}) and (\ref{eq:12}) we see that the interaction is repulsive and there are no
bound states; see, however, Section~C below.

\bigskip

{\bf B.} \quad The gauge theoretic model of Ref.~\cite{ref:1} 
 fails to describe particles with arbitrary statistics.  We now
give a quantum field theory that does the job, but it is not a gauge theory.

The quantum mechanical description is well known \cite{ref:5a}: it makes use of the
scale-invariant
${1 / x^2}$ potential, which may also be represented with the help of an exchange
operator $R$ in a ``covariant" derivative \cite{ref:5}.  The two-body equation reads
\begin{equation}
i \hbar\partial_t \psi (t; x_1, x_2) = {\hbar^2 \over 2m} \left\{ \Bigl( {1 \over i}
\partial_{x_1} +  {i\nu \over x_1-x_2} R \Bigr)^2 +
\Bigl( {1 \over i} \partial_{x_2} + {i \nu \over x_2-x_1} R \Bigr)^2 \right\}
\psi (t; x_1, x_2)
\label{eq:23}
\end{equation}
where $R f(x_1, x_2) = f(x_2, x_1)$.  Hence acting on even wave functions (\ref{eq:23})
gives
\begin{equation}
i\hbar \partial_t \psi = {\hbar^2 \over 2m} \Bigl( - \partial_{x_1}^2 - \partial_{x_2}^2 +
{2 \nu (\nu-1) \over (x_1 - x_2)^2} \Bigr) \psi .
\label{eq:24}
\end{equation}
(A similar expression is gotten with odd wave functions.) 
Note that
the
``covariant" derivatives in (\ref{eq:23})
commute, but the interaction cannot
be removed, since it is equivalent to (\ref{eq:24}).

For a field theoretical description of the many-body problem, we use a Lagrangian and
Hamiltonian as in (\ref{eq:5}), (\ref{eq:6}) with
\begin{equation}
\Pi (t,x) = \Bigl( \partial_x + \nu \int  {\rm d}y   {1 \over x-y} \rho(t,y)
\Bigr)
\Psi (t,x)
\label{eq:25}
\end{equation}
This is {\it not\/} a gauge-covariant derivative; being real, $\int {\rm d}y\, {\rho (t,y) /
(x-y)}
$ is {\it not\/} a gauge connection.  Note the Hamiltonian (\ref{eq:6}) is
{\it not\/} normal ordered, but one can reorder $H$: with $\Pi$ given by (\ref{eq:25}), one
finds
\begin{equation}
H= {\hbar^2 \over 2m} \Bigl(\int {\rm d} x \,\Psi^{* \prime} \Psi' + \nu  (\nu -1)
\int {\rm d}x\, {\rm d} y {: \rho (t,x) \rho (t,y) : \over (x-y)^2}\Bigr) .
\label{eq:26}
\end{equation}
(The term involving six $\Psi$ fields vanishes by symmetry.)  This is precisely the field
theoretic description of (\ref{eq:24}).
\bigskip

{\bf C.} \quad We now return to the model of Ref.~\cite{ref:1} and examine it as a
classical field theory.  The equation of motion that follows from
(\ref{eq:5})--(\ref{eq:7}) reads
\begin{equation}
 i\hbar \partial_t \Psi = - {\hbar^2 \over 2m} (\partial_x \pm i\varkappa^2 \rho)^2 \Psi
\pm \hbar \varkappa^2 j \Psi
\label{eq:30}
\end{equation}
where the current $j$ is given by 
\begin{equation}
j = {\hbar \over 2im} \Bigl( \Psi^* (\partial_x \pm i \varkappa^2 \rho) \Psi -
\Psi (\partial_{\varkappa} \mp i \varkappa^2 \rho) \Psi^* \Bigr)
\label{eq:31}
\end{equation}
and satisfies a continuity equation with $\rho$:
\begin{equation}
\dot{\rho} + j' = 0 .
\label{eq:32}
\end{equation}
Next we redefine the $\Psi$ field by 
\begin{equation}
\Psi (t,x) = e^{\mp i \varkappa^2 \int^x {\rm d}y \rho (t,y)} \psi (t,x)
\label{eq:33}
\end{equation}
(the lower limit is immaterial, as it affects only the phase of $\psi$) so that
\begin{equation}
i \hbar \partial_t \psi (t,x) \pm\hbar \varkappa^2 \int^x {\rm d}y \dot{\rho} (t,y) \psi (t,x)
= - {\hbar^2 \over 2m} \psi^{''} (t,x) \pm \hbar \varkappa^2 j \psi (t,x).
\label{eq:34}
\end{equation}
But the integral on the left side may be evaluated with the help of (\ref{eq:32}), and
taken to the right, leading to our final equation, which is a Schr\"odinger equation with a
{\it current\/}  ($j$) nonlinearity: 
\begin{eqnarray}
 i \hbar\partial_t \psi &=& - {\hbar^2 \over 2m}
 \psi'' \pm 2 \hbar \varkappa^2 j \psi\label{eq:35} \\
j &=& {\hbar \over 2im} \left( \psi^* \psi' - \psi\psi^{* \prime} \right) . \nonumber
\end{eqnarray}
This is to be contrasted with the familiar nonlinear Schr\"odinger equation, where the
nonlinearity involves the {\it charge density\/} ($\rho$):
\begin{equation}
 i \hbar\partial_t \psi = -{\hbar^2 \over 2m} \psi'' - \lambda \rho \psi .
\label{eq:36}
\end{equation}
Our equation (\ref{eq:35}) is similar to ``derivative nonlinear Schr\"odinger equations"
\cite{ref:7a}.

Influenced by the known solutions to (\ref{eq:36}), we seek a one-soliton solution to
(\ref{eq:35}) of the form
\begin{equation}
\psi = e^{-i (\omega t - kx)} \sqrt\rho .
\label{eq:37}
\end{equation}
 With (\ref{eq:37})
\begin{equation}
j = v \rho \ \ , \qquad v \equiv {\hbar\, k \over m}
\label{eq:38}
\end{equation}
so that our  equation (\ref{eq:35}) coincides with (\ref{eq:36})  when 
$\lambda$ is set equal to $\mp 2 \hbar\varkappa^2 v$.  For positive $\lambda$,
(\ref{eq:36}) possesses a single-soliton solution.  In our case $\mp 2 \hbar \varkappa^2$
always has a definite sign, depending on the initial choice of sign in (\ref{eq:3}). 
Then (\ref{eq:35}) also
possesses a single-soliton solution, provided $v$ is chosen so that $\mp
2\hbar\varkappa^2v$ is positive.  It is seen that the soliton of the Schr\"odinger equation
(\ref{eq:35}), with a current nonlinearity, always moves in one direction (determined by
the sign of $v$) -- it is a chiral soliton. This is in contrast to the usual Schr\"odinger
equation (\ref{eq:36}), with a charge-density nonlinearity, whose solitons can move on a
line in both directions. 
 
Henceforth, for definiteness, we take the lower sign and $\varkappa>0$; then the soliton
solution exists for positive $v$, and takes the profile
\begin{equation}
\psi_{\rm soliton} = e^{i ({mv/ \hbar}) (x-ut)}\frac1\varkappa \sqrt{\hbar\over
2m  v} \, {\alpha \over \cosh \alpha (x-vt)} .
\label{eq:39}
\end{equation}
Here $u\equiv\omega / k$ and $\alpha^2 \equiv \bigl(m^2 v^2 / \hbar^2\bigr) \left(1 -
2u/ v
\right) >0 $, which is required to be positive, {\it i.e.}, $u<v/2$.

The soliton's dynamical parameters may be evaluated as follows.  Setting
\begin{mathletters}
\label{eq:allof40}
\begin{eqnarray}
N &=& \int {\rm d}x\, \rho  \label{eq:40a} \\
\noalign{\noindent {we find}}
N_{\rm soliton} &=& {\hbar \alpha \over  \varkappa^2 m v} =
\frac1{\varkappa^2} (1-2u/v)^{1/2} . \label{eq:40b}
\end{eqnarray}
\end{mathletters}
The field energy, in terms of the rephased field $\psi$ is 
\begin{mathletters}
\label{eq:41}
\begin{eqnarray}
E &=& {\hbar^2 \over 2m} \int {\rm d}x \psi^{*\prime} \psi'  \label{eq:41a} \\
\noalign{\noindent and with the solution (\ref{eq:39}) one has}
E_{\rm soliton} &=& \half M v^2 \label{eq:41b}
\end{eqnarray}
\end{mathletters}
where
\begin{equation}
M = mN (1 + \third \varkappa^4 N^2) .
\label{eq:42}
\end{equation}
The field momentum has an unconventional form, owing to the lack of Galileo invariance
\begin{mathletters}
\label{eq:42more}
\begin{eqnarray}
P &=& \int {\rm d}x (mj + \hbar\varkappa^2 \rho^2)  \label{eq:42a} \\
\noalign{\noindent giving with (\ref{eq:39})}
P_{\rm soliton} &=& Mv . \label{eq:42b}
\end{eqnarray}
\end{mathletters}
We see from these expressions, which imply
\begin{equation}
E_{\rm soliton} = \frac v2 P_{\rm soliton}
 \label{eq:36newest}
\end{equation} that
the soliton's characteristics are those of a nonrelativistic particle of mass $M$, moving with
velocity $v$ and composed of
$N$ ``constituents". While the phase velocity $u$ is arbitrary, the group velocity $v$ must
be positive and exceed $2u$. 

When $v$ is negative   there exists a ``kink" solution
\begin{equation}
\psi_{\rm kink} = e^{i (mv/\hbar) (x-ut)} \frac1\varkappa
\sqrt{\frac\hbar{2m\left|v\right|}}\,
\beta \tanh\beta(x-vt)
\label{eq:36new}
\end{equation}
where now $\beta^2 = -(m^2v^2/2\hbar^2) (1+ 2u/\left|v\right|) > 0$, which must
 be positive, {\it i.e.}, $u<v/2$.\quad  $\psi_{\rm kink}$ interpolates between the two
``vacua"
\begin{equation}
\psi_{\rm vacuum} = \pm e^{i(mv/\hbar) (x-ut)} \frac\beta\varkappa
\sqrt{\frac\hbar{2m\left|v\right|}}.
\label{eq:37new}
\end{equation}
Because $\psi_{\rm kink}$ does not fall off at large distances, the kink's dynamical
characteristics, corresponding to (\ref{eq:allof40})--(\ref{eq:42more}), diverge. But if the
kink's energy and momentum are defined by subtracting the corresponding vacuum
values, one may still establish the relation
\begin{equation}
E_{\rm kink} = \frac v2 P_{\rm kink}.
\label{eq:38new}
\end{equation}

Semiclassical quantization of our solutions remains an open problem. We note that the
solutions are neither static nor periodic, hence new quantization techniques need to be
developed. Here we observe the following. If in the quantum theory one replaces the
Hamiltonian (\ref{eq:6}) by its normal ordered version, there are no infinities in the 
consequent $N$-body quantum mechanical problem. 
The  two-body equation, for relative
motion, reads instead of (\ref{eq:10})
\begin{equation}
\Bigl( - {\hbar^2 \over m} \partial_x^2 + {P^2 \over 4m} - {P \over m} \hbar \varkappa^2
\delta (x) \Bigr) u (x) = Eu(x) .
\end{equation}
Unlike (\ref{eq:10}), this possesses a bound state, with 
\begin{equation}
E= \frac{P^2}{4m} (1 - \varkappa^4) 
\label{eq:40new}
\end{equation}
 provided $P / m$ is positive.  If ${P / m}$ is taken proportional to $v$, this condition is the
same as that for the existence of the soliton.  So we suspect that there is a relation
between the classical soliton and quantum bound states.

Justification for this conjecture may be seen from the following argument. If in
(\ref{eq:40new}) we substitute (\ref{eq:41b}) and (\ref{eq:42b}) we get
\begin{equation}
M = \frac{2m}{1-\varkappa^4} .
\label{eq:41new}
\end{equation}
On the other hand, if (\ref{eq:42}) is to be modified in the same way that one-loop
quantum effects modify formulas for the cubic Schr\"odinger equation \cite{ref:6}, we
should replace  (\ref{eq:42}) by
\begin{mathletters}
\begin{eqnarray}
M_{\rm semiclassical} &=& mN + \third m\varkappa^4(N^3-N).\label{eq:new42a}\\
\noalign{\noindent For $N=2$ this gives}
M_{\rm semiclassical} &=& 2m(1+\varkappa^4) \label{eq:new42b}
\end{eqnarray}
\end{mathletters}%
which agrees with (\ref{eq:41new}) at ``weak coupling", {\it i.e.}, small~$\varkappa$.
Although explicit solutions of the $N$-body quantum Schr\"odinger equation are not
known for $N>2$, one may establish perturbatively in $\varkappa$ that (\ref{eq:new42a})
is consistent with the quantum bound states. Details will be presented elsewhere.

\end{document}